\documentstyle[aps,epsfig,prl]{revtex}
\begin{document}
\title{ MAGNETOOPTICAL EFFECTS IN QUANTUM WELLS IRRADIATED WITH LIGHT
PULSES}
\author{D. A. Contreras-Solorio, S. T. Pavlov\cite{byline1}}
\address {Escuela de Fisica de la UAZ, Apartado Postal C-580,
98060 Zacatecas, Zac., Mexico}
\author{ L. I. Korovin, I. G. Lang}
\address{A. F. Ioffe Physical-Technical Institute, Russian
Academy of Sciences, 194021 St. Petersburg, Russia}
\twocolumn[\hsize\textwidth\columnwidth\hsize\csname
@twocolumnfalse\endcsname
\date{\today}
\maketitle \widetext
\begin{abstract}
\begin{center}
\parbox{6in} { The method of detection and investigation of the
magnetopolaron effect in the semiconductor quantum wells (QW) in a
strong magnetic field, based on pulse light irradiation and
measuring the reflected and transmitted pulses, has been proposed.
It has been  shown that a beating amplitude on the frequencies,
corresponding to the magnetopolaron energy level splitting,
 depends strongly on the exciting pulse width. The existence
of the time points of the total reflection and total
 transparency has been predicted. The high orders of the perturbation theory on
electron-electromagnetic field interaction have been taken into
account.}

\end{center}
\end{abstract}

] \narrowtext

Time resolved scattering (TRS) investigations of the excitons in
the semiconductor bulk crystals and QWs have been discussed in the
current literature\cite{1,2}. The most interesting results are due
to the discrete energy levels and pulse irradiation of the
physical subjects. It is well known also that a pair of the energy
levels close to each other results into  a new effect: The
sinusoidal beatings  on the frequency $\Delta E/\hbar$,
corresponding to the energy distance $\Delta E$ between the energy
levels, appear in the reflected pulses.

    In this paper we investigate theoretically TRS from a
semiconductor QW in a strong magnetic field (SMF). When a SMF is
directed perpendicularly to the QW surface the discrete energy
levels of the electron excitations exist there. These discrete
energy levels are characterized with the Landau and
size-quantization quantum numbers of electrons and holes. In an
ideal situation the Landau levels (LL) are equidistant. It is
clear that one cannot restrict the consideration to some pair of
the LLs and has to take into account either one LL or a large
number of the LLs. The last variant has been used in
Ref.~\onlinecite{3} where the ladder-type structure of the
reflected and transmitted pulses has been predicted for the
strongly non-symmetrical exciting light pulse. The ladder-type
structure is characterized with the period $2\pi/\omega_{\mu}$,
where $\omega_{\mu}=|e|H/\mu c$ is the cyclotron frequency,  $\mu
=\mu_e\mu_h/(\mu_e+\mu_h),  m_e (m_h)$ is the electron (hole)
effective mass, $H$ is the magnetic field intensity. In the case
of arbitrary exciting pulses  the beating structure with the
frequency $\omega_\mu$
 depends on the pulse form, but generally speaking is not a
sinusoidal one.  In the case of the symmetrical exciting pulse,
the duration of which is much longer than the period
$2\pi/\omega_\mu$, the beatings on the frequency $\omega_\mu$ have
a very small amplitude.  Then one can restrict the consideration
by the only LL, with which the exciting pulse is in the resonance.

In Ref.~\onlinecite{4} the shape of the reflected and transmitted
pulses is determined in the vicinity of the resonance of the
exciting pulse carrier frequency $\omega_l$ with the only energy
level in the QW. Maybe this level is either excitonic level at
$H=0$, or a discrete energy level in the SMF. It has been shown
that under condition
\begin{equation}
\label{1} \gamma_r\ge\gamma,
\end{equation}
where $\gamma_r(\gamma)$ is the inverse radiative (non-radiative)
lifetime of the electronic excitation, the strong change of the
transmitted and reflected pulses comparatively with the shape of
the exciting pulse  has to happen.

The above-mentioned regards to the case when the interaction of
the electronic excitation with other excitations in the QW (in
particular with phonons) does not essentially influence the energy
spectrum of the electronic excitations. The role of the electron
(hole) - LO phonon interaction grows sharply at the resonant
values of $H$ when the magnetopolaron effect appears\cite{5}, i.e.
electron (hole) energy level splitting happens.

The polaron state formation takes place in both 3D and 2D
semiconductor systems. In the 2D case the splitting value is about
$\alpha^{1/2}\hbar\omega_{LO}$ ~ \cite{6}, where $\alpha$ is the
Fr\"ohlich electron- LO phonon coupling constant \cite{7}. The
 usual, combined and weak magnetopolarons\cite{8} exist there. The
resonant magnetic field values for the usual magnetopolarons are
determined as
\begin{equation}
\label{2} \omega_{LO}=j\Omega_{e(h)},~~   j=1, 2, 3, \ldots
\end{equation}
where $\omega_{LO}$ is the LO phonon frequency,
$\Omega_{e(h)}=|e|H/m_{e(h)} c$ is the electron (hole) cyclotron
frequency. The fractional $j$ correspond to the weak polarons.

The values $\Delta E$ of the polaron splittings for the weak
polarons are smaller than those for the usual polarons: They are
of higher order on $\alpha$ than $\alpha^{1/2}$. The values
$\gamma_{ra}$ and $\gamma_{rb}$ for both upper $(a)$ and lower
$(b)$ energy levels have been  obtained in Ref.~\onlinecite{9} as
well as the estimations of $\gamma_{a}$ and $\gamma_{b}$.

Let us show that the values $\Delta E$ of the polaron splittings
can be measured in the experiments with the pulse irradiation of a
QW in a SMF. We suppose that a light pulse incidents on the single
QW from the left perpendicular to the QW surface. The pulse
electric field intensity is
\begin {eqnarray}
\label{3} {\bf E}_0(z,t)&=&E_0{\bf e}_1
e^{-i\omega_l(t-zn/c)}\{\Theta(t-zn/c)e^{-\gamma_l(t-zn/c)/2}\nonumber\\
&+&[1-\Theta(t-zn/c)] e^{\gamma_l(t-zn/c)/2}\}+c.c.,
\end{eqnarray}
where $E_0$ is the real amplitude, ${\bf e}_l$ is the polarization
vector, $\omega_l$ is the pulse carrier frequency, $n$ is the
refraction index out the QW, $\gamma_l$ determines the pulse
duration, $\Theta(x)$ is the Haeviside step function. After the
Fourier transformation we obtain
\begin {equation}
\label{4} {\bf
 E}_0(z,t)=
E_0{\bf e}_1\int_{-\infty}^{\infty}d\omega e^{-i\omega(t-zn/c)}
D_0(\omega)+c.c.,
\end{equation}
\begin {equation}
\label{5} D_0(\omega)={i\over
2\pi}[(\omega-\omega_l+i\gamma_l/2)^{-1}-
(\omega-\omega_l-i\gamma_l/2)^{-1}].
\end{equation}
When $\gamma_l\rightarrow 0$ the pulse (3) transits into a
monochromatic wave. Let us suppose that the incident wave has a
circular polarization. We imply that both waves of the circular
polarization  correspond to the excitations of  two types of the
EHP, the energies of which are equal. Let us consider a QW, the
width of which is much smaller than the light wave  length value
$c/\omega_l n$. Then the electric fields on the left (on the
right) of the QW are determined by the expressions~\cite{4}
\begin{equation}
\label{6} {\bf
 E}_{left(right)}(z,t)={\bf E}_0(z,t)+\Delta{\bf E}_{left(right)},
\end{equation}
\begin {eqnarray}
\label{7} \Delta{\bf E}_{left(right)}(z,t)=E_0{\bf
e}_l\times\nonumber\\ \times\int_{-\infty}^{\infty}d\omega
e^{-i\omega(t\pm zn/c)} D(\omega)+c.c.,
\end{eqnarray}
where the upper (lower) sign corresponds to the index  $ <left>
(<right>)$,
\begin {equation}
\label{8}
D(\omega)=-4\pi\chi(\omega)D_0(\omega)/(1+4\pi\chi(\omega)),
\end{equation}
\begin{eqnarray}
\label{9} \chi(\omega)=(i/4\pi)\sum_\rho
(\gamma_{r\rho}/2)[(\omega-\omega_{\rho}+i\gamma_{\rho}/2)^{-1}\nonumber\\
+(\omega+\omega_{\rho}+i\gamma_{\rho}/2)^{-1}],
\end{eqnarray}
where $\rho$ is the index of the excited  state,
$\hbar\omega_\rho$ is the excitation energy measured from the
ground state energy. Applying Eq.~(8) means that the theory is
constructed with taking into account the highest orders of the
perturbation theory on the electron-EM field interaction
\cite{3,4,9,10,11}. The second term in the square brackets of
Eq.~(9) is non-resonant one and omitted below.

        We consider below the case of two excited energy levels
when the index $\rho$ takes two values: $\rho=1$ and
$\rho=2$.These levels are for instance the lower and upper
magnetopolaron energy levels. With the help of Eqs.~(6)-(9) we
obtain
\begin{eqnarray}
\label{10} \Delta{\bf E}_{left}(z,t)=-iE_0{\bf e}_l\nonumber\\
\times\left\{\Theta(s)\left[e^{-i\omega_ls-\gamma_ls/2}
\left({\bar{\gamma}_{r1}/2\over
\omega_l-\Omega_1+i(G_1-\gamma_l)/2}\right.\right.\right.\nonumber\\
\left.+{\bar{\gamma}_{r2}/2\over
\omega_l-\Omega_2+i(G_2-\gamma_l)/2}\right)\nonumber\\
-e^{-i\Omega_1s-G_1s/2}(\bar{\gamma}_{r1}/2)
\left({1\over\omega_l-\Omega_1+i(G_1-\gamma_l)/2}\right.\nonumber\\
\left.-{1\over\omega_l-\Omega_1+i(G_1+\gamma_l)/2}\right)\nonumber\\
-e^{-i\Omega_2s-G_2s/2}(\bar{\gamma}_{r2}/2)
\left({1\over\omega_l-\Omega_2+i(G_2-\gamma_l)/2}\right.\nonumber\\
\left.\left.-{1\over\omega_l-\Omega_2+i(G_2+\gamma_l)/2}
\right)\right]\nonumber\\
+(1-\Theta(s))e^{-i\omega_ls+\gamma_ls/2}
\left({\bar{\gamma}_{r1}/2\over
\omega_l-\Omega_1+i(G_1+\gamma_l)/2}\right.\nonumber\\
\left.\left.+{\bar{\gamma}_{r2}/2\over
\omega_l-\Omega_2+i(G_2+\gamma_l)/2)}\right)\right\},
\end{eqnarray}
where $s=t+zn/c$. The expression for $\Delta E_{right}(z,t)$
differs from Eq.~(10) by  the substitution $p=t-zn/c$ instead of
$s$. The designations are used
\begin{eqnarray}
\label{11} (\Omega-iG/2)_{1(2)}={1\over 2}
\Big[\tilde{\omega}_1+\tilde{\omega}_2\nonumber\\
\left.\pm\sqrt{(\tilde{\omega}_1-\tilde{\omega}_2)^2
-\gamma_{r1}\gamma_{r2}}\right];\nonumber\\
 \tilde{\omega}_{1(2)}=\omega_{1(2)}-i\Gamma_{1(2)}/2,~~
\Gamma_{1(2)}=\gamma_{1(2)}+\gamma_{r1(2)};\nonumber\\
 \bar{\gamma}_{r1}=\gamma_{r1}+\Delta\gamma,~~
 \bar{\gamma}_{r2}=\gamma_{r2}-\Delta\gamma;\nonumber\\
\Delta\gamma=\{\gamma_{r1}[\Omega_2-\omega_2-i(G_2-\gamma_2)/2]\nonumber\\+
\gamma_{r2}[\Omega_1-\omega_1-i(G_1-\gamma_1)/2]\}\nonumber\\
\times[\Omega_1-\Omega_2+i(G_2-G_1)/2]^{-1}.
\end{eqnarray}
The upper (lower) sign in Eq.~(11) corresponds to 1(2). The values
$\Omega_{1(2)}$ and $G_{1(2)}$ are real by their definition. At
$\gamma_l=0$ the expression Eq.~(10) transits into the formula for
the induced field at a monochromatic irradiation\cite{9}. At
$\gamma_{r2}=0$ we obtain from Eq.~(10)
\begin{eqnarray}
\label{12} \Delta{\bf E}_{left}(z, t)=-iE_0{\bf
e}_l(\gamma_{r1}/2)\nonumber\\
 \times\left\{\Theta(s)
\left[{e^{-i\omega_ls-\gamma_ls/2}\over
\omega_l-\omega_1+i(\Gamma_1-\gamma_l)/2}\right.\right.\nonumber\\
-e^{-i\omega_ls-\Gamma_1s/2}
\left({1\over\omega_l-\omega_1+i(\Gamma_1-\gamma_l)/2}\right.\nonumber\\
\left.\left.-{1\over\omega_l-\omega_1+i(\Gamma_1+\gamma_l)/2}\right)
\right]\nonumber\\
\left.+(1-\Theta(s)){e^{-i\omega_ls+\gamma_ls/2}\over
\omega_l-\omega_1+i(\Gamma_1+\gamma_l)/2}\right\}\nonumber\\+c.c.,
\end{eqnarray}
which corresponds to the case of the single excited energy level.
Comparing Eq.~(10) and Eq.~(12) one finds that the fields from two
levels not only add, but the renormalization of the frequencies
$\omega_{1(2)}$,
 broadenings $\Gamma_{1(2)}$ and the factors $\gamma_{r1(2)}$
has happened.
 They are substituted by $\Omega_{1(2)}, G_{1(2)}$ and
 $\bar{\gamma}_{r1(2)}$, respectively. In the case of two merging
levels, i.e. under condition
$\gamma_{r1}=\gamma_{r2},~\gamma_{1}=\gamma_{2},~
\omega_{1}=\omega_{2}, $ we obtain from Eq.~(10) an expression of
type of Eq.~(12) in which the value $\gamma_{r1}$ has to be
substituted by $2\gamma_{r1}$. It means that in the case of the
twice degenerated level, the twofold value of the inverse
radiative lifetime  figures in all the formulae.

        In Figs.~1-5 the modulus ${\cal P}$ and ${\cal T}$ of the exciting and transiting
pulses as the functions of $p=t-zn/c$ , the module ${\cal R}$ of
the reflected pulse flux as function of $s=t+zn/c$ and the
absorption which, is defined at $z=0$ as
\begin{equation}
\label{13} {\cal A}={\cal P}-{\cal T}-{\cal R},
\end{equation}
are represented. The modulus of the energy fluxes are represented
in units $(c/2\pi n)E_0^2$.

        It follows from Eq.~(10) that the results for the energy fluxes
are dependent on the parameters $\omega_l-\omega_1$ and
$\gamma_l$, characterizing the exciting pulse, and the parameters
$\Delta\omega=\omega_1-\omega_2,~\gamma_{r1},~ \gamma_{r2},~
\gamma_1,~ \gamma_2$, characterizing the system with two excited
levels. It follows from Eq.~ (10) that there are resonances on the
frequencies $\omega_l=\Omega_1,~ \omega_l=\Omega_2$ and beatings
on the three frequencies $\omega_l-\Omega_1,~\omega_l-\Omega_2,~
\Omega_1-\Omega_2$ exist.

In the case
\begin{equation}
\label{14} \omega_l=\Omega_1,~~
\gamma_{r1}=\gamma_{r2}=\gamma_{r},~~ \gamma_1=\gamma_2=\gamma
\end{equation}
the results for the fluxes are represented in Figs.~1-5. Thus,
only parameters $\gamma_l,~\Delta\omega,~ \gamma_r,~ \gamma$ are
variable ones. Obviously, under conditions Eq.~(14) the
 beatings are possible only at the
frequency $\Delta\Omega=\Omega_1-\Omega_2$, if the value
$\Delta\Omega\not=0$ , which does not always fulfilled as we will
see below. Figs. 1, 2 correspond to the parameters
\begin{eqnarray}
\label{15} \Delta\omega=0.00665,~~
\gamma_{r}=0.00005,~~\gamma=0.0005,\nonumber\\
\gamma_{l}=0.0005~(Fig. 1),~~ \gamma_l=0.005~(Fig. 2).
\end{eqnarray}
We use the arbitrary units because the results for the fluxes
depend on the parameter interrelations  only. If the eVs are used
then $\Delta\omega$  from Eq.~(15) correspond [8,9] to the usual
polaron at $j=1$ (see Eq.~(2)) at the QW width $d=300 A$ for GaAs.
The relation $\gamma/\gamma_r=10$ is chosen arbitrarily.
$\Omega_1\simeq\omega_1$ and $\Omega_2\simeq\omega_2$ when the
parameters of Eq.~(15) are used.

        Fig. 1 and Fig. 2 differ from each other sharply.
The beatings are almost invisible in Fig. 1. It means that these
results correspond to the induced electric field Eq.~(12) under
condition $\omega_l=\omega_1$. In Fig. 2 the beatings at the
frequency $\Delta\omega=\omega_1-\omega_2$ are seen brightly in
the reflected pulse. Reflection and absorption are much smaller
than unity in Fig.~ 1, 2, and reflection is much smaller than
absorption. These features characterize the case
$\gamma_r<<\gamma$. As for beatings, their appearance depends
sharply on the pulse duration $\gamma_l^{-1}$. Beatings are
clearly visible for the short pulses when
$\gamma_l\simeq\Delta\omega$. This is clear physically: The
frequencies, which are close to $\omega_1$ and to $\omega_2$, are
essential in the frequency spectrum Eq.~(5) of the short pulse.

        The former parameters $\Delta\omega=0.00665,~
\gamma_r=0.00005,~$ but $\gamma=0$ correspond to Fig. 3 and Fig.
4. The parameter $\gamma_l$ in Fig. 3 and Fig. 4 coincide with
$\gamma_l$ in Fig. 1 and Fig. 2, respectively. The beatings in the
reflected and absorbed pulses are expressed only in Fig. 4 when
$\gamma_l\simeq\Delta\omega$. In Figs.~ 3, 4 reflection and
absorption are prolonged more than in the Figs.~ 1, 2, because in
the case $\gamma=0$ the prolongation is determined by the value
$\gamma_r^{-1}$.

In Figs. 2, 3, 4 there is the time point $t_0$, where absorption
equals to zero and changes its sign. At $s_0=t_0$ the reflected
flux is equal to the exciting flux and the transmitted flux equals
to zero at $p_0=t_0$. This is the time point in which the field
from the right of the QW equals zero (the total reflection point).
An analysis shows that in the case of a single excited level at
$\omega_l=\omega_1$ incidentally the especial point exists only at
$t_0>0$, which corresponds to the condition $\gamma_l>\gamma,$
\begin{equation}
\label{16}
t_0={2\over\Gamma_1-\gamma_l}\ln{2\gamma_{r1}\gamma_l\over
(\gamma_l-\gamma_1)(\Gamma_1+\gamma_l)}.
\end{equation}

 In Figs.~ 1, 3 the
curves are presented at which the level 2 almost does not
practically influence. The condition $\gamma_l>\gamma$ is
satisfied for Fig.~ 3, but not for Fig.~ 1. Therefore the point of
total reflection is presented in Fig.~ 3 and it is absent in Fig.~
1. In Ref.~\onlinecite{4} it has been shown that the total
reflection points exist also for other forms of pulses including
non-symmetrical ones.

        The relations $${\cal P}\simeq 0,~ {\cal T}\simeq
{\cal R},~ {\cal A}\simeq -2{\cal T}$$ are fulfilled for the times
very much larger than $\gamma_l^{-1}$ in Figs.~ 2, 3, 4. The sense
of them is clear: If $\gamma_l>>\gamma+\gamma_r$ then the fields
created by the exciting pulse  are very small at quite large
values $t\pm zn/c$, and the fluxes of ${\cal T}$ and ${\cal R}$
are determined only by the induced fields $\Delta
E_{left(right)}$. The picture becomes a symmetrical one relative
to the plane $z=0$ where the QW is placed, the fluxes from the
left and right of the plane are equal in absolute values. The
absorbed flux is negative and equals to the sum of fluxes going to
the left and to the right. That means that the QW gives back the
accumulated energy.

        In Fig.~ 5 the parameters
\begin{eqnarray}
\label{17} \Delta\omega=0.00665,\nonumber\\
\gamma_{r}=0.00666,~~\gamma=0.0001 ,~~\gamma_{l}=0.001
\end{eqnarray}
are used, i. e.
 the special case $\Delta\omega\simeq\gamma_r$ is represented. The
beating absence is due to $\Delta\Omega=0$, but not to the small
influence of one of the levels. Indeed, as it follows from Eqs.~
(11), (14), $\Omega_1=\Omega_2=(\omega_1+\omega_2)/2$ at
$|\Delta\omega|\le\gamma_r$. Thus, if $\omega_l=\Omega_1$, the
pulse carrier frequency is situated exactly between the two levels
in the case of Fig.~ 5.

The beatings are absent in Fig. 5, but the figure structure
differs radically from the structure for the case of two energy
levels. In Fig. 5 there is the time point $t_x$ of the total
transparency in which reflection and absorption are equal to 0 and
the transmitted flux is equal to the exciting flux. An analysis
shows that under conditions Eq.~ (14) and
$|\Delta\omega|=\gamma_r$
\begin{equation}
\label{18}
t_x={2\over\gamma_l-\Gamma}\ln{(\gamma_l-\gamma)(\Gamma+\gamma_l)^2\over
2\gamma_l(\gamma_l^2+\gamma_r^2-\gamma^2)},
\end{equation}
and the total transparency point exists in the case
$\gamma_l<\Gamma$, which is fulfilled for Fig.~ 5. The electric
field to the left of the QW equals zero in the time point $t_x$.

        In all the figures, besides Fig. 1, it is seen that the absorbed
flux equals zero in the total reflection or total transparency
point and changes its sign, i.e. becomes negative one. That means
that the electronic system at first accumulates the energy of the
created electron-hole pairs and then gives it back. In the case
$\gamma_1=\gamma_2=0$ all the accumulated energy comes back as
radiation, i.e. the integral absorption equals zero. This result
is true for the cases of Figs.~ 3, 4.

        In Figs. 1- 5 there are only a few examples of the shapes of the
reflected, absorbed and transmitted pulses, corresponding to some
combinations from the seven parameters. It is worth stressing that
the experimental results on the pulse irradiation of the QWs in a
magnetic field allow us in principle to detect and investigate the
magnetopolaron states. Measuring of time lags of the reflected and
transmitted pulses allows us to determine the lifetimes
 of the polaron states. The special subject of interest is
the investigation of the points of the total reflection and total
transition, the first appear in the case of short pulses. For the
usual polarons the case of Figs. 1, 2 is the most real one. With
the help of the pulse irradiation of the QWs one can investigate
not only usual, but weak polarons also. The smaller $\Delta\omega$
for the weak polarons are even preferable because, first, the
beating frequency diminishes and, second, one can use more longer
pulses, than in the case of usual polarons.

\section{Acknowledgements}
        S.T.P thanks the Zacatecas Autonomous University and the National
Council of Science and Technology (CONACyT) of Mexico for the
financial support and hospitality. D.A.C.S. thanks CONACyT
(27736-E) for the financial support.
       Authors are grateful
 to A. D'Amor for a critical reading of the
manuscript.
       This work has been partially supported by the Russian
Foundation for Basic Research and by the Program "Solid State
Nanostructures Physics".

%
\begin{figure}
 \caption{ The
modulus of the exciting, transmitted, reflected and absorbed
energy fluxes as time functions: $ \omega_l=\Omega_1,~
\Delta\omega=0.00665,~\gamma_r=0.00005,~\gamma=0.0005,~\gamma_l=0.0005.$
All the parameters are given in  eV.}
\end{figure}
\begin{figure}
 \caption{ Same
as Fig~1 for $\gamma_l=0.005.$}
\end{figure}
\begin{figure}
\caption{Same as Fig.~1 for $\gamma=0,~\gamma_l=0.0005.$}
\end{figure}
\begin{figure}
 \caption{Same
as Fig~1 for $\gamma=0,~\gamma_l=0.005.$}
\end{figure}
\begin{figure}
\caption{Same as Fig.~1 for $\gamma_r=0.00666,~\gamma=0.0001,
~\gamma_l=0.001.$}
\end{figure}

\end{document}